\RequirePackage{tikz}

\documentclass[sn-basic, iicol]{sn-jnl}
\pdfoutput=1

\RequirePackage{natbib}
\setcitestyle{aysep={}} 

\usepackage{program}
\usepackage{amssymb}
\usepackage{threeparttable}
\usepackage{changepage}
\usepackage{booktabs}
\usepackage{hhline}
\usepackage{multirow}
\usepackage{breqn}
\usepackage{enumitem}  
\usepackage{amsthm}
\usepackage{algorithm}
\usepackage{algpseudocode}
\usepackage{geometry}
\usepackage{mathtools}
\usepackage{graphicx}
\usepackage{subfig}
\usepackage{tabularx}
\usepackage{tabularx}
\usepackage{adjustbox}
\usepackage{import}
\usepackage{float}
\usepackage{pgfplots}

\usepgfplotslibrary{groupplots}
\pgfplotsset{compat=newest}

\usepackage[labelsep=space]{caption}

\jyear{2022}%

\theoremstyle{thmstyleone}%
\theoremstyle{thmstyletwo}%

\theoremstyle{thmstylethree}%

\begin{document}

\title[Paper's title]{Survival analysis for user disengagement prediction: question-and-answering communities' case}


\author{\fnm{Hassan} \sur{Abedi Firouzjaei}}\email{hassan.abedi@ntnu.no}



\affil{\orgdiv{Department of Computer Science}, \orgname{Norwegian University of Science and Technology (NTNU)}, \orgaddress{
\city{Trondheim}, 
\country{Norway}}}




\abstract{We used survival analysis to model user disengagement in three distinct questions-and-answering communities in this work. We used the complete historical data of \{Politics, Data Science, Computer Science\} Stack Exchange communities from their inception until May 2021, which include the information about all users who were members of one of these three communities. Furthermore, formulating the user disengagement prediction as a survival analysis task, we utilised two survival analysis techniques to model and predict the probabilities of members of each community becoming disengaged. Our main finding is that the likelihood of users with even a few contributions staying active is noticeably higher than the users who were making no contributions; this distinction may widen as time passes. Moreover, the results of our experiments indicate that users with more favourable views towards the content shared on the platform may stay engaged longer. Finally, the observed pattern holds for all three communities, regardless of their themes.}




\maketitle

\section{Introduction}\label{sec1}

Online question-and-answering (QA) social networks\footnote{In this work, we use the terms question-and-answering platform, social network, and community interchangeably} like Stack Overflow and Quora are dependent on their users' contributions for proper functioning. Arguably, the main functionality of a QA platform is to connect two types of users~\citep{Kuzmeski2009}; on one side, people who seek answers for their questions and on the other side, people who are willing to share their knowledge and expertise with others. Nevertheless, a user who joined and made many contributions to the community may become uninterested and then disengaged after a while. By disengaged, we mean the situation where users---as individuals who previously made contributions (eg answered questions and participated in debates)---suddenly stopped their activities (ie, there is no sign of them even visiting the platform's web pages).
Moreover, it is not known whether these users left the community or not, but they did not perform any activity on the platform for a relatively long period of time (eg more than a year). In this context, disengagement might have happened for various reasons; eg it might have occurred because disengaged users thought the platform had an elitist or even toxic culture. Another reason could have been that user interests changed drastically over time, and the platform hosting the QA community could not adapt to the change in an agile way.

At the very least, a high disengagement rate has adverse effects on the overall quality of the service of a QA social network and platform. For example, suppose all the experts (ie, users who post answers perceived as high quality by the community) become disengaged within a few months of joining and being active. In that case, the quality of answers might plummet, which may increase the rate of users being disengaged from the community. In the worst-case scenario, one could expect the situation where the QA platform loses the bulk of its contributors, which in turn would lead to its demise.

Survival analysis~\citep{Cox2018} is a family of statistical methods and techniques that can help model and predict the time of the occurrence of an event of interest. Initially, it emerged out of the field of medical research to find the probability of a patient surviving a disease such as cancer---hence the term \textit{survival analysis}. More recently, survival analysis methods have found widespread use in new areas such as customer churn analysis~\citep{Dias2020, Rothmeier2021} and credit risk scoring~\citep{Stepanova2002}, mainly due to their flexibility and power in accurately and reliably modelling the problems posed in these areas. 

\begin{figure}[h]
\centerline{\includegraphics[scale=0.45]{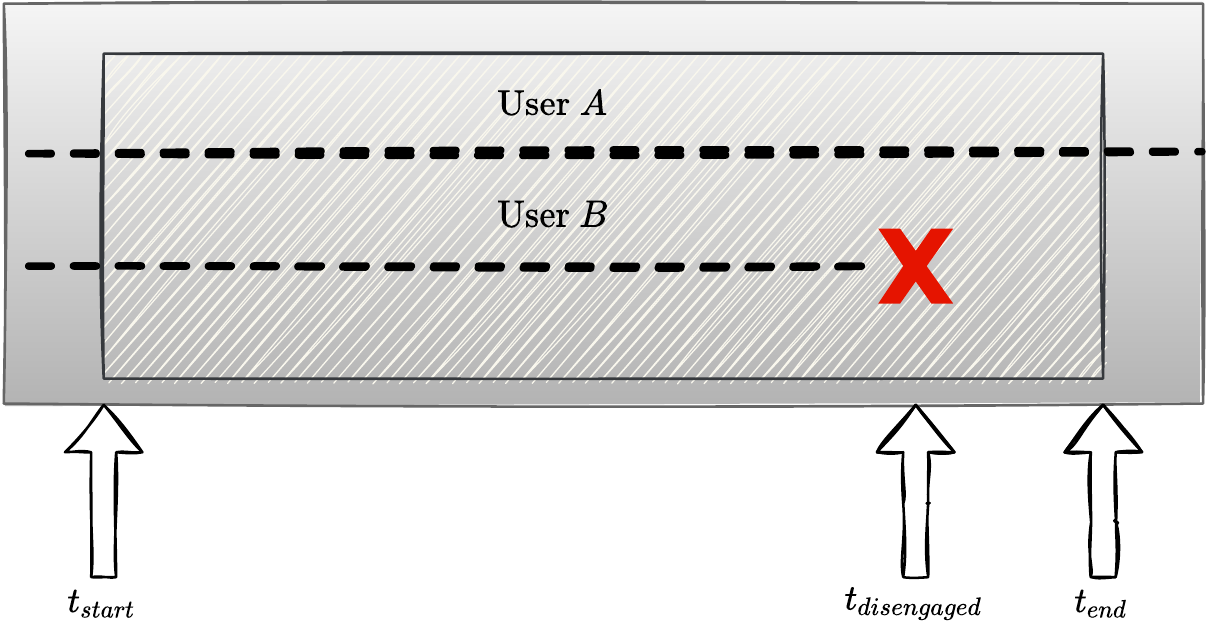}}
\caption{User $A$ and user $B$ joined the platform in the past; during a period of observation which started at $t_{start}$ and ended at $t_{end}$, $B$ became disengaged at $t_{disengaged}$. $A$ did not become disengaged during the observation, but it is not known that he will become disengaged in the future or not; information about $A$'s disengagement is censored} 
\label{fig:eksempel}
\end{figure}

In this work, we used survival analysis to study user disengagement in three distinct QA social networks, namely, \{Politics, Data Science, Computer Science\} Stack Exchange. To our knowledge, this is the first work that applied survival analysis to quantify and study user disengagement using the entire historical data of online QA social networks. Fig.~\ref{fig:eksempel} illustrates how disengagement prediction can be seen and formulated as a survival analysis task.

Following are the main contributions of our work:

\begin{itemize}
    \item We study the factors likely to be associated with the probability that users of QA communities will stay active for an extended period. 
    For the first time, we analyse the relationships between attributes related to users' contributions and their engagement time.
    \item We propose to exploit behavioural, and content-based user attributes to estimate the engagement time on three comprehensive datasets from distinct QA communities. 
\end{itemize}

The rest of this article is organised as follows. Section~\ref{relwks} discusses the related work. Section~\ref{pream} presents preliminary concepts related to survival analysis and introduces techniques used to model and evaluate the problem of user disengagement prediction. Section~\ref{data} gives an overview of the dataset and the methodology used to represent users and the engagement time. Section~\ref{results} presents the results of the experiments and Section~\ref{disc} discusses the results. Section~\ref{limit} discusses the limitations of our work, and gives an outline for direction of future work. Finally, Section~\ref{conc} concludes the paper.

\section{Related work}\label{relwks}

QA platforms like Stack Exchange and Quora provide an accessible environment for knowledge sharing. Due to the importance of this role which earlier was played by mailing lists, newsgroups and IRC channels, the interest in studying phenomena in these platforms exploded lately. For example, \citet{Joyce2006} studied the continued user participation in newsgroups. \citet{Guan2018} investigated the factors related to users motivation to participate in community activities, especially knowledge contribution. \citet{Jin2015} studied the elements that were influencing user knowledge contributions in QA platforms, incorporating three theories of social capital theory, social exchange theory, and social cognitive theory in their work.

Furthermore, the use of survival analysis methods is also getting popularity where an analogy could be made between the problem and the task of survival analysis~\citep{Wang2019}.
\citet{Yang2010} used survival analysis methods to analyse and study user retention in three QA communities. Although their work is similar to our work, we used the data for the whole lifespan of the communities, where their work mainly focused on a limited period. \citet{Dupret2013} used survival analysis to analyse the user engagement in a dataset of questions and answers from Yahoo Answers based on user absence time. Most works in this area that are related to data from QA communities are mainly focused on the data from a few larger communities such as Stack Overflow~\citep{Ortega2014}.

\section{Preamble}\label{pream}

\subsection{Survival analysis}
Survival analysis or time-to-event analysis~\citep{Cox2018} is a set of statistical models and methods for estimating the time it takes for a particular event of interest to happen. In a typical survival analysis task, a group of individuals (eg patients) are observed for a period. For each individual, the time when the event of interest happened is recorded. Usually, the event will not occur for all the individuals in the period of observation. The situation when the event of interest did not happen for an individual during the observation is called \textit{censoring}. The goal of survival analysis is to find the probability of happening of the event of interest. In this regard, survival analysis is similar to regression analysis but with a major difference, where survival analysis models take into account the information related to individuals for whom the event did not take place, ie, the censored individuals. This difference allows for obtaining more accurate estimations. Although survival analysis originated from the field of medical research, mainly for estimating the time a patient would live after being diagnosed having a deadly disease such as breast cancer, it has gained much attention in other areas such as customer churn analysis and prediction~\citep{Dias2020} and time to occurrence of a fault in a system~\citep{Widodo2011}.

Formally, $T \geq 0$ is a random variable that models the time for an event of interest to happen; $f(t)$ and $F(t)$ are its probability distribution and cumulative probability distribution, respectively.

\begin{equation}
    F(t) = \int_{-\infty}^t \! f(x) \, \mathrm{d}x
\end{equation}

Furthermore, $S(t)$, called the survival function, is defined as the probability that the event did not happen before time $t$. (Typically, when $S(t)$ is plotted, it is called the survival curve.)

\begin{equation}
    S(t) = P[T > t] = 1 - F(t)
\end{equation}

The hazard function $h(t)$, is the instantaneous occurrence rate of the event of interest, and is defined as:

\begin{equation}
    h(t) = \lim_{\mathrm{d}t \to 0} {P[t \leq T < T+\mathrm{d}t | T \geq t] \over \mathrm{d}t} = {f(t) \over S(t)} 
\end{equation}

Survival and hazard functions can be connected via the following formula:

\begin{equation}
    S(t) = e^{-\int_{0}^t \! h(x) \, \mathrm{d}x}
\end{equation}

Given $n$ individual samples, each sample $i \in [1...n]$ is represented as triplet $(A^i, E^i, T^i)$ where:

\begin{itemize}
    \item $A^i \in R^d$ is a $d$-dimensional real-valued vector of individual features (ie, user attributes in our context);
    \item $E^i \in \{0, 1\}$ is the variable indicating the event of interest happened when $E^i=1$ or not (censored) when $E^i=0$, for individual $i$ during the observation;
    \item $T^i=min(t_i, t_{end})$ is the time when the event happened for individual $i$ during the observation period; $t_{end}$ is the time when the observation was ended. $T^i = t_{end}$ (ie, event did not happen) indicates sample $i$ is censored.
\end{itemize}

The main task of the survival analysis methods is to estimate $h(t)$ and $S(t)$.

\subsection{Kaplan-Meier estimator}

Kaplan-Meier estimator~\citep{Kaplan1958} is a non-parametric model that calculates the survival function $\hat{S}_{KM}(t)$ of a homogeneous cohort, ie, the individuals in the same cohort (or group) share the same survival function. Given $N$ individual samples in a cohort, it assumes that there are $J$ distinct actual event times such that $t_1 < t_2 <...<t_{J}$ when $J \leq N$, then:

\begin{equation}
    \hat{S}_{KM}(t) = \prod_{t_j \leq t}^{} (1 - {d_j \over n_j}),
\end{equation}

where $d_j$ is the individuals who experienced an event and $n_j$ is the number of individuals that did not experience the event in time interval $[t_{j-1}, t_j]$. 

Kaplan-Meier method only uses the information from $E^i$ and $T^i$ to estimate the survival function.

\subsection{Random survival forests}

\citet{Ishwaran2008} proposed the random survival forests (RSF) model, which is an extension to the random forests ensemble model~\citep{Breiman2001} for working with censored data. The general idea for creating an RSF model for a particular dataset is as follows~\citep{Utkin2019, Ishwaran2008}:

\begin{enumerate}
    \item Bootstrap $q$ samples from the data, where $q$ is the number of trees. On average, each sample excludes 37\% of the original data as out-of-bag (OOB) data.
    \item Grow a survival tree for each bootstrap sample. At each node of the tree, select $\sqrt{m}$ (ie, a subset of variables used during the node split) candidate variables. Then split the node using the variable that maximises the survival difference between its children nodes.
    \item Furthermore, grow the tree to be full under the constraint where no leaf node should have less than $d > 0$ deaths. The value of $d$ is a hyperparameter, similar to $q$, which is chosen to produce the best results.
    \item Compute the cumulative hazard function (or the survival function) for each tree.
    \item Use the OOB data to calculate the prediction error for the ensemble cumulative hazard function (or the survival function).
\end{enumerate}

Different implementations of RSF mainly differ in their splitting rule. Ideally, the splitting rule should maximise the survival difference across two dataset partitions. In this paper, we used the implementation from PySurvvial library~\citep{pysurvival_cite}.

\subsection{Concordance index}

The concordance index (or C-index for short) is a generalisation of the area under the ROC curve (AUC), which supports censored data~\citep{Harrell1982}. C-index widely is used as an evaluation metric of the performance of survival models. It summarises the model's discriminatory power, which is how well a model can rank the survival times of samples. Similar to AUC, the value of the C-index ranges from 0.5 to 1, where 1 indicates the best performance.

More formally, given $S(t)$ be the survival function estimated by some survival model, let $t_1^*,...,t_s*$ be a set of fixed time points, eg $t_1,...,t_N$ where $N$ is a distinct time index. Then C-index is defined as:

\begin{equation}
    C = {1 \over M} \sum_{i:E^i=1}^{}\sum_{j:t_i<t_j}^{}{\textbf{1}[S(t_i^*) > S(t_j^*)]},
\end{equation}

where $M$ is the total number of comparable pairs and $\textbf{1}[.]$ is a function that will return 1 if its argument is true or 0 otherwise. Note that there are slightly different definitions for C-index in other works. In this work, we used the definition proposed by~\citet{Utkin2019}.

\subsection{Log-rank test}

The log-rank test~\citep{Mantel1966} is a non-parametric statistical test for comparing the hazard functions, ie, $h(t)$, of two cohorts/groups of individuals. The null hypothesis is that the hazard functions of two groups, eg group $1$ and $2$, are equal, ie, $h_1(t) = h_2(t)$. The Log-rank test assumes that survival probabilities (ie, the probabilities of not becoming disengaged in our context) stay the same over time. It is widely used to check whether the underlying survival distributions of two groups are the same or are different, essentially. 

\section{Data}\label{data}

\subsection{Data description}

As mentioned earlier, we used data from three online QA platforms that are \{Politics (Pol), Data Science (DS), Computer Science (CS)\} Stack Exchange (SE). Pol SE is an ad-hoc QA community focused on politically-themed content, such as questions related to the nature of democracy and the state of human rights. DS SE covers topics concerning the widespread field of data science. And CS SE covers topics related to computer science. We chose these three communities for two reasons. Firstly, although the sizes of these communities are smaller than the sizes of some other QA communities hosted on SE like Stack Overflow, nevertheless, the chosen communities are thriving in their niche. Secondly, each of these communities is more or less focused on separate fields that, although they might share some topics, are different enough to be viewed as distinct. It allows us to search for possible patterns related to disengagement, regardless of the specific topics of a field.

The datasets of the three communities were downloaded from the Stack Exchange data dump available on~\href{Archive.org}{https://archive.org/download/stackexchange}~\footnote{The data are available under Creative Commons licences}. The data included the complete historical information about the questions and answers posted on the three QA communities from their inception until May 2021. Table~\ref{tab:desc} shows the general information about the datasets.

\begin{table}
\begin{center}
\caption{Information about the datasets}
  \begin{tabular}{lccc}
    \toprule 
    \multirow{2}{*}{{Characteristic}} &
    \multicolumn{3}{c}{{Dataset}} \\ &
    \scriptsize{{Pol}} & \scriptsize{{DS}} & \scriptsize{{CS}} \\ \hline
    \midrule
    \scriptsize{Number of questions} & \scriptsize{12416} & \scriptsize{28950} & \scriptsize{40792} \\ \hline
    \scriptsize{Number of users} & \scriptsize{31,242} & \scriptsize{100582} & \scriptsize{113434}\\ \hline
    \scriptsize{Number of answers} & \scriptsize{25909} & \scriptsize{32334} & \scriptsize{46785} \\ \hline
    \scriptsize{Number of comments} & \scriptsize{135648} & \scriptsize{64244} & \scriptsize{167038} \\ \hline
     \scriptsize{Year community founded} & \scriptsize{2012} & \scriptsize{2014} & \scriptsize{2008} \\ \hline
    \bottomrule \hline
  \end{tabular}
\label{tab:desc}
\end{center}
\end{table}

\subsection{User attributes}

The bulk of users in QA platforms do not make any contributions. These users, who are referred to as lurkers in some previous work~\citep[eg][]{Tagarelli2018}, can be differentiated from the normal users, who include the experts, by their level of contribution to the platform. In this work, two categories of user attributes were used in order to investigate the relationship between the user contributions and the probability of disengagement. Namely, \textit{behavioural attributes} and \textit{content-based attributes}.

\subsubsection{Behavioural attributes}

We identified five user attributes in the datasets that directly correspond to the level of user contribution. These attributes primarily are based on the information related to user behaviour that seems crucial to the proper functioning of the platform. Table~\ref{tab:userattrsb} includes the name and description of these attributes.

\subsubsection{Content-based attributes}

In addition to behavioural attributes, we picked up a set of content-based user attributes. These attributes hint at how the contributions made by each user might have been perceived favourably by the community, ie, other users. The primary motivation is that users can make indirect contributions to the platform, eg asking a question that starts a stream of debates over a controversial topic such as \textit{refugee crisis} in the context Pol SE community. And the information about this type of indirect user contribution, which is not only limited to the behaviour of a particular user, can be extracted and utilised from user content (eg mainly from metadata of users' posts). Table~\ref{tab:userattrscb} includes the name and description of the content-based attributes employed in this work.

\subsection{User representation}

Based on the two types of user attributes mentioned earlier, we constructed a numerical vector called \textit{user representation vector} (URV) for each user. URV of user $i$ is defined as:

\begin{equation}
    URV_{i} = (A_1^i,A_2^i,...A_j^i)
\end{equation}

Where $A_j^i$ is the corresponding user attribute from Tables~\ref{tab:userattrsb} and~\ref{tab:userattrscb}.

\begin{table}[h]
\caption{Information about behavioural user attributes}
\begin{center}
\begin{tabular}{c p{4.5cm}}
\toprule
{User Attribute}      
& {Description} \\ \hline  \midrule
\scriptsize{$A_1^i$} & \scriptsize{Number of downvotes cast by user $i$}\\ \hline
\scriptsize{$A_2^i$} & \scriptsize{Number of upvotes cast by user $i$}\\ \hline
\scriptsize{$A_3^i$} & \scriptsize{Number of questions posted by user $i$}\\ \hline
\scriptsize{$A_4^i$} & \scriptsize{Number of answers posted by user $i$}\\ \hline
\scriptsize{$A_5^i$} & \scriptsize{Number of comments written by user $i$}\\ \hline
\bottomrule
\end{tabular}
\end{center}
\label{tab:userattrsb}
\end{table}

\begin{table}[h]
\caption{Information related to the content-based attributes}
\begin{center}
\begin{tabular}{c p{4.5cm}}
\toprule
{User Attribute}      
& {Description} \\ \hline  \midrule
\scriptsize{$A_6^i$} & \scriptsize{Average number of times questions posted by user $i$ were viewed}\\ \hline
\scriptsize{$A_7^i$} & \scriptsize{Average number of comments written for questions posted by user $i$}\\ \hline
\scriptsize{$A_8^i$} & \scriptsize{Average score (ie, sum of upvotes and downvotes given to question) of the questions posted by user $i$}\\ \hline
\scriptsize{$A_9^i$} & \scriptsize{Average score (ie, sum of upvotes and downvotes given to answer) of the answers posted by user $i$}\\ \hline
\scriptsize{$A_{10}^i$} & \scriptsize{Average number of comments written for answers posted by user $i$}\\ \hline
\scriptsize{$A_{11}^i$} & \scriptsize{Average number of upvotes given to the comments posted by user $i$}\\ \hline
\bottomrule
\end{tabular}
\end{center}
\label{tab:userattrscb}
\end{table}

\subsection{Dichotomizing users}

Moreover, we dichotomised users into pairs of disjoint sets (or groups) using each one of behavioural user attributes. The main idea is that users can be partitioned into two groups naturally, where the criterion for the split is whether a user has made a particular type of contribution or not. Categorising users in two disjoint sets based on the value of a behavioural attribute allowed us to investigate the importance of one specific user attribute, eg by comparing the survival curves of two disjoint sets of users who posted at least one question and who did not. Table~\ref{tab:usercharsets} includes the information about each group of users and its counterpart.
\begin{table}[h]
\caption{Sets of users defined by dichotomizing users based on their behavioural attributes corresponding to their contribution}
\begin{center}
\begin{tabular}{c p{5cm}}
\toprule
{User Set} & {Definition} \\ \hline \midrule

\scriptsize{$Q$} & \scriptsize{Users who posted at least one question}\\
\scriptsize{$\overline{Q}$} & \scriptsize{Users who posted no question}\\\hline

\scriptsize{$A$} & \scriptsize{Users who answered at least one question}\\
\scriptsize{$\overline{A}$} & \scriptsize{Users who did not answer any question}\\\hline

\scriptsize{$C$} & \scriptsize{Users who commented on at least one question/answer}\\
\scriptsize{$\overline{C}$} & \scriptsize{Users who did not answer any question}\\\hline

\scriptsize{$U$} & \scriptsize{Users who upvoted at least one question/answer/comment}\\
\scriptsize{$\overline{U}$} & \scriptsize{Users who did not upvoted} \\\hline

\scriptsize{$D$} & \scriptsize{Users who downvoted at least one question/answer} \\
$\overline{D}$ & \scriptsize{Users who did not downvote}\\\hline

\bottomrule
\end{tabular}
\end{center}
\label{tab:usercharsets}
\end{table}

\subsection{Disengagement criterion}

What amounts to the event of a user becoming disengaged is domain-dependent and thus can vary in different settings. For example, normally, in medical research, the event usually is the patient's death~\citep{Cox2018}. In this work, suitable to our need, we opted to use the information about the last time a user visited the QA platform to detect disengagement. The information about the last time a user visited is available in the \textit{LastAccessDate} column from the \textit{Users} table in each dataset. The activity time of a user was calculated based on the difference in the number of months since the user joined the platform until the last recorded activity time of the user (ie, \textit{LastAccessDate} associated with the user). For user $i$, if the number of months since his last visit to the platform exceeded a certain threshold value, he would be tagged as a disengaged user (ie, $E^i = 1$); otherwise, the user's state was considered still active (ie, $E^i = 0$) which means the information about user's disengagement was censored. More precisely, let $t^i_{l}$ be the last time user $i$ been seen visiting the platform, and $\theta$ be the threshold value, then the value of $E^i$ would be set based on the following relation:

\begin{equation}
E^i = f_{\theta}(t^i_l)= 
\begin{dcases}
    0,& \text{if } \hat{d}(t^i_l, t_{d}) \leq \theta\\
    1,              & \text{otherwise}
\end{dcases}
\end{equation}

where $t_{d}$ is the last recorded time in the dataset, and $\hat{d}(t^i_l, t_{d})$ is the time difference between $t_l^i$ and $t_{d}$ in months.
In this work, we used two threshold values (ie, $\theta$) of 24 and 36 months. Subsequently, users who had not visited the platform for more than two and three years were considered disengaged.

\section{Results}\label{results}

\subsection{Results from Kaplan-Meier}

Kaplan-Meier method was used to estimate (within the confidence interval of 95\%) the survival functions (ie, $S(t)$) of sets of users dichotomised based on the definitions shown in Table~\ref{tab:usercharsets}.
The implementation from Lifeline library~\citep{DavidsonPilon2021} was used to produce the survival curves.
Figs.~\ref{fig:scp},~\ref{fig:scds}, and~\ref{fig:sccs} show the survival curve of each pair sets of users for \{Pol, DS, CS\} SE datasets respectively. Table~\ref{tab:kmusersets} includes the information about the size and proportion of each user set dichotomised based on a single behavioural attribute. As mentioned earlier, for each user, the label indicating whether the user is disengaged or not (ie, $E^i$) were censored if the difference between the user's last activity time and $t_{d}$ was less than or equal to 24 and 36 months, respectively. Log-rank test (with $p < 0.005$) was performed on each pair of curves.

\begin{table}
\begin{center}
\caption{Sizes of each user set per dataset}
  \begin{tabular}{cccc}
    \toprule 
    \multirow{2}{*}{User Set} &
    \multicolumn{3}{c}{Dataset} \\
    & \scriptsize{Pol} & \scriptsize{DS} & \scriptsize{CS} \\ \hline
    \midrule
    
    \scriptsize{$|Q|$} & \scriptsize{3775 (12\%)} & \scriptsize{16041 (16\%)} & \scriptsize{20841 (18\%)} \\
    \scriptsize{$|\overline{Q}|$} & \scriptsize{27466 (88\%)} & \scriptsize{84540 (84\%)} & \scriptsize{92592 (82\%)} \\ \hline 
    
    \scriptsize{$|A|$} & \scriptsize{3743 (12\%)} & \scriptsize{7226 (7\%)} & \scriptsize{7020 (6\%)} \\
    \scriptsize{$|\overline{A}|$} & \scriptsize{27498 (88\%)} & \scriptsize{93355 (93\%)} & \scriptsize{106413 (94\%)} \\ \hline 
    \scriptsize{$|C|$} & \scriptsize{6358 (20\%)} & \scriptsize{12056 (12\%)} & \scriptsize{16788 (15\%)} \\
    \scriptsize{$|\overline{C}|$} & \scriptsize{24883 (80\%)} & \scriptsize{88525 (88\%)} & \scriptsize{96645 (85\%)} \\ \hline 
    
    \scriptsize{$|U|$} & \scriptsize{11025  (35\%)} & \scriptsize{16328 (16\%)} & \scriptsize{20767 (18\%)} \\
    \scriptsize{$|\overline{U}|$} & \scriptsize{20216 (65\%)} & \scriptsize{84253 (84\%)} & \scriptsize{92666 (82\%)} \\ \hline 
    
    \scriptsize{$|D|$} & \scriptsize{1596 (5\%)} & \scriptsize{732 (1\%)} & \scriptsize{1143 (1\%)} \\
    \scriptsize{$|\overline{D}|$} & \scriptsize{29645 (95\%)} & \scriptsize{99849 (99\%)} & \scriptsize{112290 (99\%)} \\ \hline
    
    \bottomrule \hline
  \end{tabular}
\label{tab:kmusersets}
\end{center}
\end{table}

\begin{figure*}[ht]

\begin{center}
    \subfloat[ {$\theta=24$}]{%
    \resizebox{0.25\textwidth}{!}{%
      \input{figures/tikz/p/24/survival_curve_questions_24.tikz}}}
    \subfloat[ {$\theta=36$}]{%
    \resizebox{0.25\textwidth}{!}{%
      \input{figures/tikz/p/36/survial_curve_questions_36.tikz}}}
    \subfloat[ {$\theta=24$}]{%
    \resizebox{0.25\textwidth}{!}{%
      \input{figures/tikz/p/24/survial_curve_answers_24.tikz}}}
    \subfloat[ {$\theta=36$}]{%
    \resizebox{0.25\textwidth}{!}{%
      \input{figures/tikz/p/36/survial_curve_answers_36.tikz}}}\\
      
    \subfloat[ {$\theta=24$}]{%
    \resizebox{0.25\textwidth}{!}{%
      \input{figures/tikz/p/24/survial_curve_comments_24.tikz}}}
    \subfloat[ {$\theta=36$}]{%
    \resizebox{0.25\textwidth}{!}{%
      \input{figures/tikz/p/36/survial_curve_comments_36.tikz}}}
    \subfloat[ {$\theta=24$}]{%
    \resizebox{0.25\textwidth}{!}{%
      \input{figures/tikz/p/24/survial_curve_upvotes_24.tikz}}}
    \subfloat[ {$\theta=36$}]{%
    \resizebox{0.25\textwidth}{!}{%
      \input{figures/tikz/p/36/survial_curve_upvotes_36.tikz}}}\\
      
    \subfloat[ {$\theta=24$}]{%
    \resizebox{0.25\textwidth}{!}{%
      \input{figures/tikz/p/24/survial_curve_downvotes_24.tikz}}}
    \subfloat[ {$\theta=36$}]{%
    \resizebox{0.25\textwidth}{!}{%
      \input{figures/tikz/p/36/survial_curve_downvotes_36.tikz}}} \\

\caption{Survival curves for Pol SE dataset estimated using Kaplan-Meier method}
\label{fig:scp}
\end{center}
\end{figure*}

\begin{figure*}[h!]

\begin{center}
    \subfloat[ {$\theta=24$}]{%
    \resizebox{0.25\textwidth}{!}{%
      \input{figures/tikz/ds/24/survial_curve_questions_24.tikz}}}
    \subfloat[ {$\theta=36$}]{%
    \resizebox{0.25\textwidth}{!}{%
      \input{figures/tikz/ds/36/survial_curve_questions_36.tikz}}}
    \subfloat[ {$\theta=24$}]{%
    \resizebox{0.25\textwidth}{!}{%
      \input{figures/tikz/ds/24/survial_curve_answers_24.tikz}}}
    \subfloat[ {$\theta=36$}]{%
    \resizebox{0.25\textwidth}{!}{%
      \input{figures/tikz/ds/36/survial_curve_answers_36.tikz}}}\\
      
    \subfloat[ {$\theta=24$}]{%
    \resizebox{0.25\textwidth}{!}{%
      \input{figures/tikz/ds/24/survial_curve_comments_24.tikz}}}
    \subfloat[ {$\theta=36$}]{%
    \resizebox{0.25\textwidth}{!}{%
      \input{figures/tikz/ds/36/survial_curve_comments_36.tikz}}}
    \subfloat[ {$\theta=24$}]{%
    \resizebox{0.25\textwidth}{!}{%
      \input{figures/tikz/ds/24/survial_curve_upvotes_24.tikz}}}
    \subfloat[ {$\theta=36$}]{%
    \resizebox{0.25\textwidth}{!}{%
      \input{figures/tikz/ds/36/survial_curve_upvotes_36.tikz}}}\\
      
    \subfloat[ {$\theta=24$}]{%
    \resizebox{0.25\textwidth}{!}{%
      \input{figures/tikz/ds/24/survial_curve_downvotes_24.tikz}}}
    \subfloat[ {$\theta=36$}]{%
    \resizebox{0.25\textwidth}{!}{%
      \input{figures/tikz/ds/36/survial_curve_downvotes_36.tikz}}} \\
\end{center}

\caption{Survival curves for DS SE dataset estimated using Kaplan-Meier method}
\label{fig:scds}
\end{figure*}

\begin{figure*}[h!]

\begin{center}
    \subfloat[ {$\theta=24$}]{%
    \resizebox{0.25\textwidth}{!}{%
      \input{figures/tikz/cs/24/survial_curve_questions_24.tikz}}}
    \subfloat[ {$\theta=36$}]{%
    \resizebox{0.25\textwidth}{!}{%
      \input{figures/tikz/cs/36/survial_curve_questions_36.tikz}}}
    \subfloat[ {$\theta=24$}]{%
    \resizebox{0.25\textwidth}{!}{%
      \input{figures/tikz/cs/24/survial_curve_answers_24.tikz}}}
    \subfloat[ {$\theta=36$}]{%
    \resizebox{0.25\textwidth}{!}{%
      \input{figures/tikz/cs/36/survial_curve_answers_36.tikz}}} \\
      
    \subfloat[ {$\theta=24$}]{%
    \resizebox{0.25\textwidth}{!}{%
      \input{figures/tikz/cs/24/survial_curve_comments_24.tikz}}}
    \subfloat[ {$\theta=36$}]{%
    \resizebox{0.25\textwidth}{!}{%
      \input{figures/tikz/cs/36/survial_curve_comments_36.tikz}}}
    \subfloat[ {$\theta=24$}]{%
    \resizebox{0.25\textwidth}{!}{%
      \input{figures/tikz/cs/24/survial_curve_upvotes_24.tikz}}}
    \subfloat[ {$\theta=36$}]{%
    \resizebox{0.25\textwidth}{!}{%
      \input{figures/tikz/cs/36/survial_curve_upvotes_36.tikz}}} \\
      
    \subfloat[ {$\theta=24$}]{%
    \resizebox{0.25\textwidth}{!}{%
      \input{figures/tikz/cs/24/survial_curve_downvotes_24.tikz}}}
    \subfloat[ {$\theta=36$}]{%
    \resizebox{0.25\textwidth}{!}{%
      \input{figures/tikz/cs/36/survial_curve_downvotes_36.tikz}}} \\
\end{center}

\caption{Survival curves for CS SE dataset estimated using Kaplan-Meier method}
\label{fig:sccs}
\end{figure*}

\subsection{Results from RSF}

We used k-fold cross-validation (with k=5) in 30 runs to make the predictions (using RSF models). The values of model hyperparameters such as the number of trees (ie, $q$) have been tested in order to choose the ones that lead to the best results. We used C-index to evaluate the performance of the models. 
To train and evaluate the RSF-based models, only data for users who had some contributions were used. In other words, only the information of users belonging to $Q \cup A \cup C \cup U \cup D$ (from Table~\ref{tab:usercharsets}) was used.
For each user, three URVs were constructed, using the behavioural user attributes only, content-based attributes only, and finally, the combination of both. Table~\ref{tab:resrsf} includes the average C-indexes computed for the RSF models over the runs. 

\begin{table*}[h]
\centering
\caption{Average C-index for RSF models using different attribute sets; higher C-index indicates better prediction}
\begin{tabular}{llllll}
   \hline \toprule
   \multirow{2}{*}{{Dataset}} & 
   \multicolumn{1}{c}{} & \multicolumn{2}{c}{$\theta = 24$} & \multicolumn{2}{c}{$\theta = 36$} \\ \cmidrule(lr){3-4} \cmidrule(lr){5-6}
   & {Attributes} & Mean & STD & Mean & STD \\
   \hline
\midrule
& Behavioural only & 0.75 & 0.01 & 0.76 & 0.01 \\ 
Pol & Content-based only & 0.68 & 0.01 & 0.68 & 0.01 \\ 
& Behavioural plus content-based & 0.75 & 0.01 & 0.76 & 0.01 \\ \hline
& Behavioural only & 0.66 & 0.01 & 0.66 & 0.01 \\ 
DS & Content-based only & 0.61 & 0.01 & 0.63 & 0.01 \\ 
& Behavioural plus content-based & 0.68 & 0.01 & 0.70 & 0.01 \\ \hline
& Behavioural only & 0.68 & 0.00 & 0.68 & 0.01 \\ 
CS & Content-based only & 0.62 & 0.01 & 0.63 & 0.01 \\ 
& Behavioural plus content-based & 0.69 & 0.01 & 0.68 & 0.01 \\
\hline
\bottomrule
\end{tabular}
\label{tab:resrsf}
\end{table*}

\subsection{Attribute importance}

\begin{figure*}[h!]
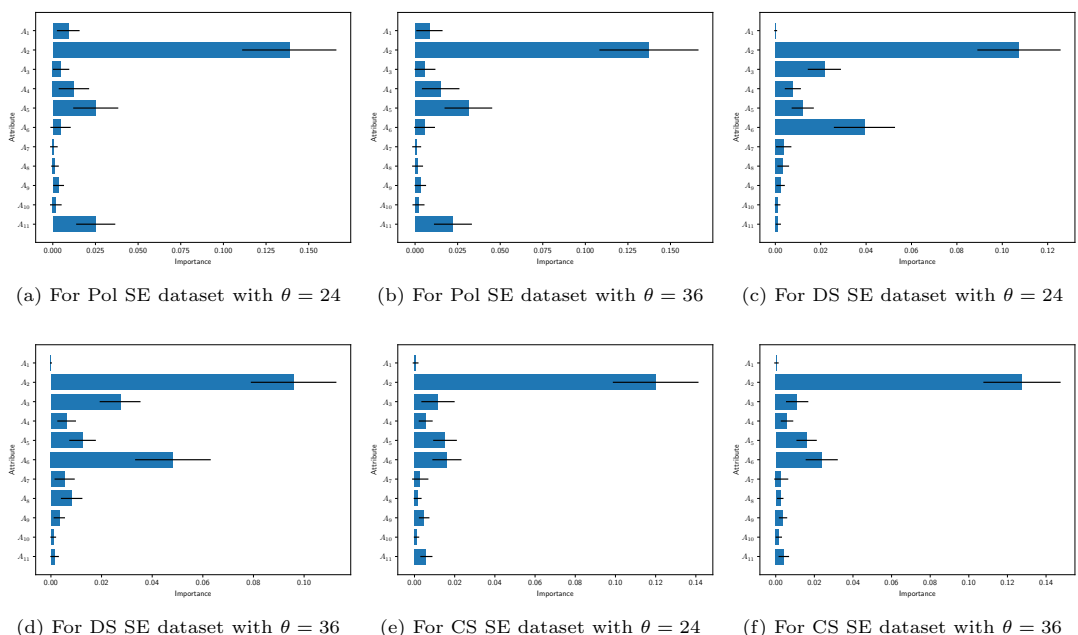

\begin{center}
    \subfloat[{For Pol SE dataset with $\theta=24$}]{%
    \resizebox{0.3\textwidth}{!}{%
    \import{figures/pgf/p}{bar_plot_with_error_bars_24_cb+b.pgf}}}
    \subfloat[{For Pol SE dataset with $\theta=36$}]{%
    \resizebox{0.3\textwidth}{!}{%
    \import{figures/pgf/p}{bar_plot_with_error_bars_36_cb+b.pgf}}}
    \subfloat[{For DS SE dataset with $\theta=24$}]{%
    \resizebox{0.3\textwidth}{!}{%
    \import{figures/pgf/ds}{bar_plot_with_error_bars_24_cb+b.pgf}}}
    
    \subfloat[{For DS SE dataset with $\theta=36$}]{%
    \resizebox{0.3\textwidth}{!}{%
    \import{figures/pgf/ds}{bar_plot_with_error_bars_36_cb+b.pgf}}}
    \subfloat[{For CS SE dataset with $\theta=24$}]{%
    \resizebox{0.3\textwidth}{!}{%
    \import{figures/pgf/cs}{bar_plot_with_error_bars_24_cb+b.pgf}}}
    \subfloat[{For CS SE dataset with $\theta=36$}]{%
    \resizebox{0.3\textwidth}{!}{%
    \import{figures/pgf/cs}{bar_plot_with_error_bars_36_cb+b.pgf}}}
\end{center}
\caption{Average permutation importance of each attribute; models are trained and evaluated using behavioural and content-based attributes simultaneously over the datasets shown in Table~\ref{tab:desc}}
\label{fig:attimp}
\end{figure*}

We used the permutation importance measure~\citep{Breiman2001, Molnar2020} present in RSF models to rank each user attribute. The attributes which permuting their values caused more significant prediction errors were ranked more important. We chose permutation importance mainly because of its intuitive definition and subsequent interpretation, which is based on the idea that the importance of a variable is the increase in model error when the variable's information is destroyed via value permutation~\citep{Molnar2020}. Moreover, it provides a compact global insight into the behaviour of the model. Fig.~\ref{fig:attimp} shows the per-dataset average permutation importance of each attribute on the prediction results of the RSF models used in this work.

\section{Discussion}\label{disc}

Based on results from the Kaplan-Meier method, we observed: i) the underlying hazard function of each set of users seem to be different; ii) the probability that users with even a few contributions (eg user asked one question) are noticeably higher than other users who did not contribute to the platform. We observed a distinctive difference between the survival functions of the users who contributed to the platform and those who did not contribute. This pattern, which is present in all three datasets regardless of the community niche, confirms the finding from previous related studies such as the ones reported in~\citep{Joyce2006, Yang2010} that
suggested that users with even a few initial contributions are more likely to stay loyal than users without any contributions. The latter make the bulk of the users. Furthermore, the gap between the probability of disengagement of two groups seems to widen over time.

Predictions using RSF models show relatively similar patterns on all three datasets. Results (shown in Table~\ref{tab:resrsf}) indicate that inclusion of the information of behavioural attributes lead to better predictions compared to the use of content-based user attributes only. Furthermore, using the mixture of the information of behavioural and content-based attributes yielded a slight improvement. The value of $\theta$ does not seem to affect the overall results.

Based on the permutation importance of attributes (see Fig.~\ref{fig:attimp}), behavioural features play a more salient role in the output of the RSF models. On average, 4 out of 5 top attributes with the most permutation importance are from behavioural user attributes.
The number of upvotes (ie, $A_2$) received the highest importance in all three datasets. Subsequently, with a noticeable difference, the average number of the times user questions were viewed (ie, $A_6$) received relatively high importance. We suspect that higher upvotes by the user might be showing that they hold a favourable view of the community (or platform) in general. On the other hand, the information about the number of downvotes did not contain much predictive information. We suspect it could be due to the small number of users with downvoting activity in the datasets.

\section{Limitations and future work}\label{limit}

There are a few limitations regarding the work done in this paper. The datasets were used only from three QA communities hosted by (the larger) Stack Exchange platform. Consequently, this work did not investigate and compare disengagement on other major platforms such as Quora. It seems interesting to compare our results with the results obtained with data from other major QA platforms in the future. Conventional assumptions related to the application of survival analysis techniques hold over our results, eg the assumption that the probabilities of disengagement of censored and none censored individuals are essentially the same. We used user inactivity for an extended period (eg two years passed since the user visited the community web pages) to distinguish between disengaged and censored users. This required the use of a time threshold in which its value is set experimentally, not based on a well-defined rule. Finally, for the majority of users, their behavioural information does not exist, which makes it hard to investigate further the survival probabilities of users dichotomised based on the definitions given in Table~\ref{tab:usercharsets}.

It could be interesting to include the data from a more numerous and diverse set of QA platforms for future work. Furthermore, the information related to the body of the posts (eg text of questions and answers) of each user could be utilised to find the probabilities of disengagement. Additionally, methods and models that do not assume the probabilities for the disengaged and censored users are the same can be used, which theoretically should lead to better predictions.

\section{Conclusion}\label{conc}

We used survival analysis to study user disengagement using the whole historical data from three distinct QA communities from their inception up to now May 2021. We employed two categories of user attributes and investigated the importance of these attributes. Our results confirm the previous findings that users with some initial contributions (eg questions and answers) are more likely to stay active longer than a user who contributed nothing. Furthermore, based on our results, behavioural user attributes can be used to estimate the disengagement probability of each user with reasonable accuracy. 

Moreover, based on the importance of attributes used to train and evaluate the models, how favourable users see the content posted on the platform seems to affect the disengagement time.

\section*{Declarations}

\bmhead{Funding}

This work was carried out as part of the Trondheim Analytica project (please visit \url{https://www.ntnu.edu/trondheimanalytica}), supported by NTNU's Digital Transformation programme.

\bmhead{Conflict of interest}
I do not have any conflicts or competing interests to declare.

\bmhead{Availability of data and materials}
The data used in this study are publicly available from~\href{Archive.org}{https://archive.org/download/stackexchange} under Creative Commons licences. 


\bibliography{references}

\end{document}